\documentclass[11pt,twoside,a4paper]{article}
\usepackage{txfonts}
\input{epsf.tex}

\title{2mass/Hipparcos Extinctions and Distances in the Serpens -- Aquila Region\\
IV. Error Propagation. Individual stars and mean values}
\author{J. Knude\\
Niels Bohr Institute, Copenhagen University\\ 
Juliane Maries Vej 30, DK-2100 Copenhagen {\O}, Denmark\\
indus@nbi.ku.dk}
\date{March 1, 2011}
\begin{document}
\maketitle
\section{Abstract}
We estimate errors from the application of the main sequence $(H-K)_0 \ vs. \ (J-H)_0$ and $(J-K)_0 \ vs. \ M_J$ 
relations on 2mass photometry to provide stellar extinctions and distances. The error 
propagation on individual stars in the Serpens -- Aquila star forming region is investigated.
Results: errors on the $J-band$ extinction $\sigma_{A_J}$ are found in the range from 0.05 to 0.20 
mag and peak at $\approx$0.1. For the absolute $J-band$ magnitudes the $\sigma_{M_J}$ range is found 
from 0.1 to 0.5 mag and the distribution has a pronounced maximum at $\approx$0.3 mag. Individual 
relative distance errors, $\sigma_{distance}/distance$, are found in the
interval from 8$\%$ to 25$\%$ and peak at 15$\%$. If cloud distances are identified with the location
of a local maximum of the $A_J(median)/distance(median)$ variation the Serpens -- Aquila cloud may be at
203$\pm$7 pc. We propose that the median absolute deviation from the median of $A_J$ could be used as a
cloud distance indicator \\

$Keywords$: Photometry: 2mass, Astrometry: photometric parallaxes, Error distributions: 
$\sigma_{A_J}$ , $\sigma_{ M_J}$, $\sigma_{distance}$, $\sigma_{distance}/distance$, Molecular clouds: 
Serpens -- Aquila, Molecular clouds: distance  

\maketitle

\section{Introduction}

It is of some interest to investigate the precision of the extinctions and distances estimated 
from individual 2mass sources until all sky astrometry becomes available. When the photometry
is precise enough 2mass data 
might be useful to map out 3D extinction distribution in the local Milky Way. For extinction 
estimates we rely on a standard $(H-K)_0 \ vs. \ (J-H)_0$ relationship like Bessell and Brett 
\cite{BB88}. With the intrinsic colors at hand the extinction is known and
distances may subsequently be derived from a calibration of $M_J$ in terms of $(J-K)_0$.
The distance follows from $r(pc) = dex(0.2(J-M_J-A_J+5))$.
A $M_J$ calibration may be established from stars common to Hipparcos and 2mass.
If several stars are measuring a common interstellar feature, e.g. identified as a local
maximum in the line of sight median density -- distance variation, given as $A_J / distance$,
the error of the mean distances and mean extinctions pertaining to the feature might be quite small 
and perhaps even comparable to those obtained from optical photometry and trigonometric parallaxes.

\section{Intrinsic colors}

Bessell and Brett \cite{BB88} have provided  $(H-K)_0 \ vs. \ (J-H)_0$ relations for main sequence
and giant stars. The Bessell and Brett colors may be transformed to the 2mass system following
Carpenter \cite{JMC01}. New two color relations, reddening free, may be established from stars 
common to Hipparcos and 2mass where Hipparcos may provide distance selections more or less assuring 
no extinction. But such work has not been done so far. Straizys and Lazauskaite \cite{straizys09}
have recently presented intrinsic 
2mass colors for main sequence and giant stars together with spectral classes. Standard deviations 
for most of their main sequence colors are given in the range from 0.02 to 0.05 mag. 
A comparison of Straizys and Lazauskaite \cite{straizys09} main sequence to Bessell
and Brett \cite{BB88} shows a good agreement. Minor deviations are only present for early
 type stars where 
Straizys and Lazauskaite's new relation has systematically bluer $(H-K)_0$ colors than Bessell
and Brett. Since the difference is only minor and concentrated to the earlier spectral
types, which we do not use anyway, we have continued using Bessell and Brett \cite{BB88}.

\begin{figure}
\epsfxsize=8.0cm
\epsfysize=8.0cm
\epsfbox{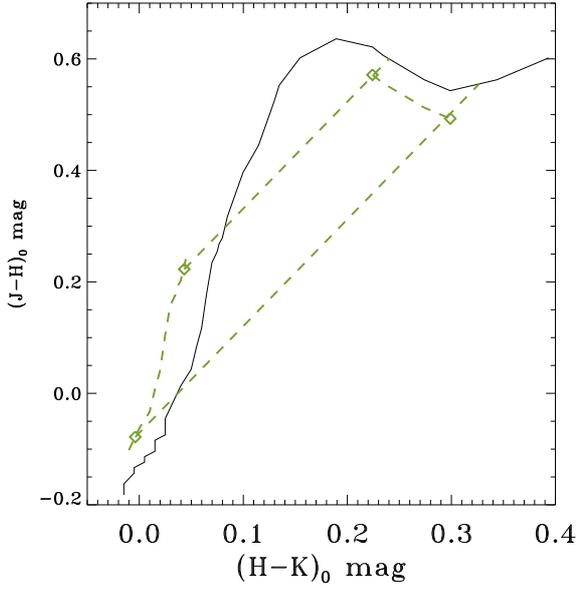} 
\caption[]{The dashed curve outlines the region of acceptance in the CCD. The solid curve is the Bessell-Brett 
\cite{BB88} standard relation for the main sequence transformed to 2mass. The upper two limitations excludes 
giant stars and little reddened M-type stars. The lower right limitation excludes early type stars where the 
slope of the standard relation approximates the reddening ratio. Stars inside the dashed curve connecting the 
four diamonds constitute the primary sample for which unique intrinsic colors may be estimated}\label{f1}
\end{figure}

For the following considerations reference is made to Fig.~\ref{f1}. The shape of the intrinsic 
CCD implies two ambiguities. For a the spectral range, approximately G4 -- M0, the main sequence 
and giant relation almost coincides. Second, the shape of the CCD signifies that beyond
about M0 it becomes a problem to decide whether a star is a much reddened early type or a less reddened
late type, unless additional luminosity information is available. As a consequence the first choice of 
reddening tracers is confined to a part of the interior of
the $(H-K)_0 \ vs. \ (J-H)_0$ relation in Fig.~\ref{f1}. We confine the observed $JHK$ sample by
the upper
straight line that has a slope equal to that of the reddening vector, $E_{J-H}$ / $E_{H-K}$,
and intersects the main sequence where the giants branch off the main sequence. The lower
confining reddening vector is introduced because the derivative of the 
$(H-K)_0 \ vs. \ (J-H)_0$ relation is almost identical to the slope of the reddening vector 
implying that the intersection, $((H-K)_0, \ (J-H)_0)$, is undetermined. 

Some stars with low or no reddening will be scattered to the blue side of the
standard relation by the photometric errors. The maximum error accepted in the present note
is 0.040 mag in each of the three 2mass bands. The curve to the blue of the standard relation
defines the blue limit for the stars included in the sample. In order to exclude most unreddened 
M-type stars we impose a upper $J-H$ boundary just below the standard relation by shifting the
relation by the maximum error, 0.040 mag, in J, H, and K. Stars inside the dashed "polygon" 
are accepted for further study. 
 The intrinsic colors are determined by translating a star in the accepted region of Fig.~\ref{f1} 
along a reddening vector to intersect the standard curve. $A_J$ is derived from the estimated
color excess $E_{J-H}$  and a standard extinction law.

\section{Absolute magnitudes}

\begin{figure}
\epsfxsize=8.0cm
\epsfysize=8.0cm
\epsfbox{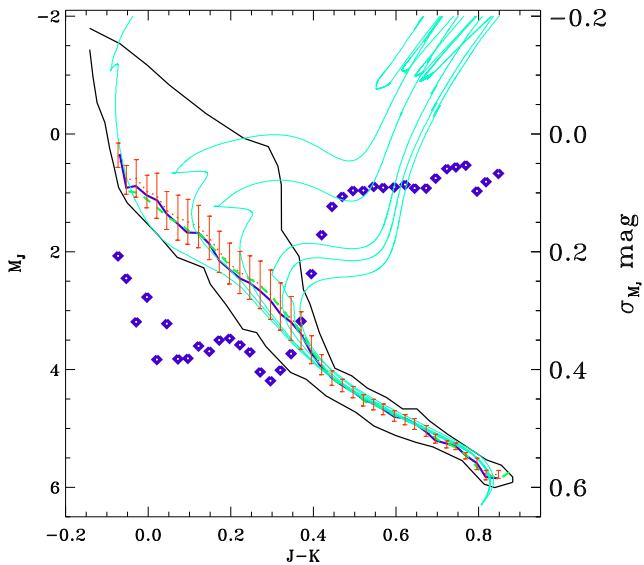} 
\caption[]{The closed solid curve is the confinement of the main sequence
sample discussed in the text and is shown together with its resulting
statistical relations calculated for 0.025 mag bins of $(J-K)$. The thick solid
curve is the median given together with the standard deviation computed for 0.050
$(J-K)$ intervals.
Isochrones from Cordier et al. \cite{CPCS07} are shown
for 0.1, 0.8, 1.5, 4.0, 5.0 and 8.0 Gyr. The diamonds show $\sigma_{M_J}$
(right hand scale) calculated for overlapping 0.050 intervals in $(J-K)$
separated by 0.025 mag and a drop from 0.4 mag to 0.1 mag is noted where the
8 Gyr isochrone turns off the main sequence}\label{f2}
\end{figure}

In order to establish the main sequence $(J-K)_0 \ vs. \ M_J$ relation we have extracted stars
common to Hipparcos, Perryman et al. \cite{PMAN97}, and 2mass, Cutri et al. \cite{cutri03}. Two
requirements are imposed on the 
astrometry. Precision: $\sigma_{\pi} / \pi \ <$ 10$\%$. Reddening free: $\pi \ >$ 10
mas $\Leftrightarrow$ a stellar distance $<$ 100 pc that commonly is accepted as a boundary
for the reddening free local volume. 

A substantial part of the Hipparcos stars has spectral and luminosity classification. From the
HR diagram of the common sample it was, however, noticed that several stars have been  misclassified
and a large fraction of the trigonometric stars does not possess a full classification. The $M_J$
calibration sample is consequently not based exclusively on fully classified stars but instead 
the main sequence is defined by the outline shown in Fig.~\ref{f2}. All stars inside this boundary, 
and which are not known to be multiple,
are accepted as main sequence stars. Median values for $(J-K)_0$ and $M_J$ are computed for 0.050 mag 
bins of  $(J-K)_0$. The resulting median $(J-K)_0 \ vs. \ M_J$ relation is plotted together with the 
errors of the means in Fig.~\ref{f2}. The scale on the right hand ordinate axis pertain to the errors 
of the mean shown as the diamonds of Fig.~\ref{f2}. The standard deviation results from photometric 
and astrometric errors, and not least from the width of the main sequence. The error of the mean is 
seen to vary from $\approx$0.4 to $\lesssim$0.1 mag, the better value is for stars redder than the 
main sequence turn off indicated by the Cordier et al. \cite{CPCS07} isochrones. Further details
are given in Knude \cite{knude10}.  

\section{Minor secondary sample}
\begin{figure}
\epsfxsize=18.0cm
\epsfysize=21.0cm
\epsfbox{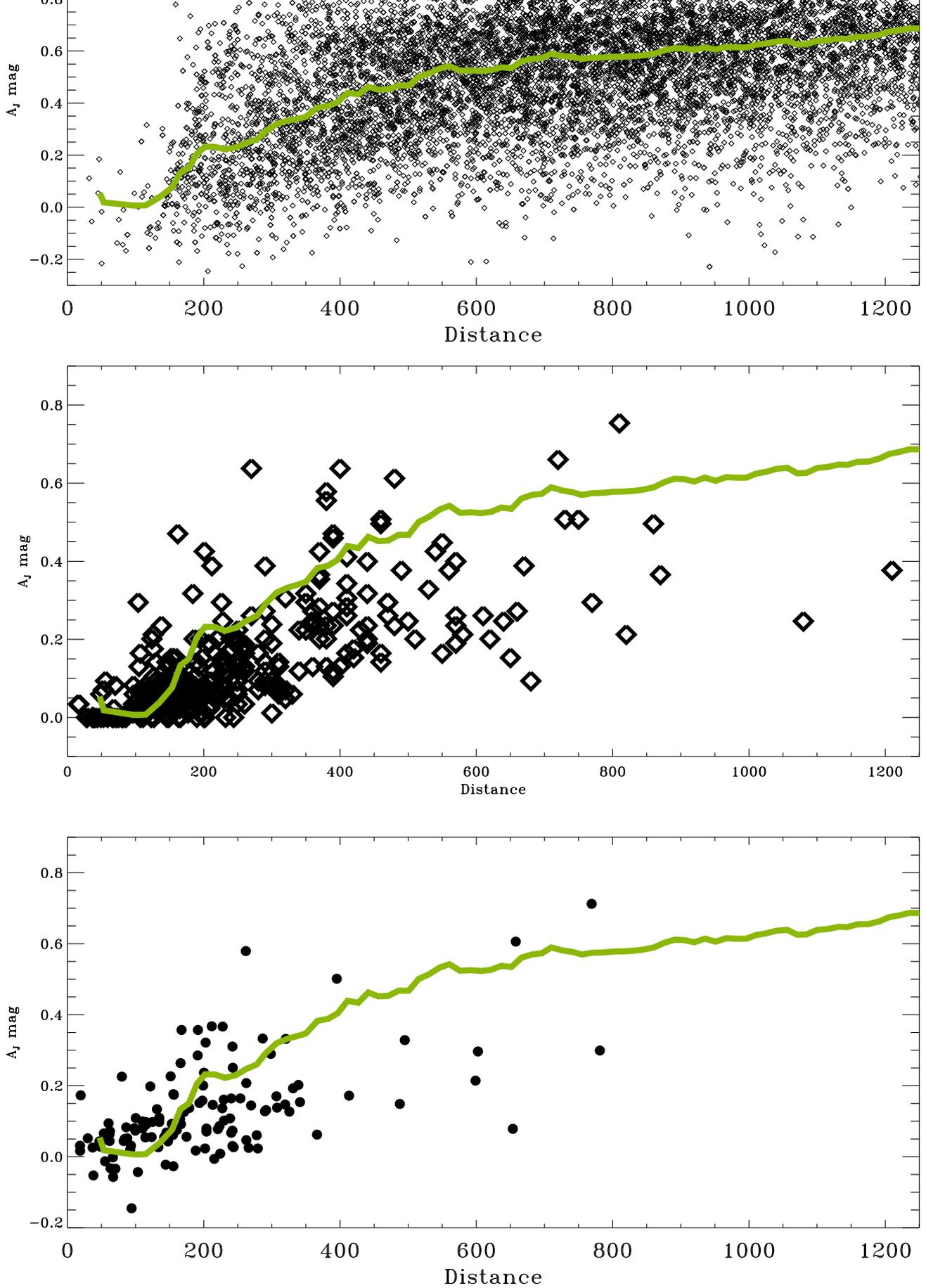} 
\caption[]{{\it Upper panel} Stars with $\sigma_{JHK} <$ 0.040 mag from the sky patch defined by
$\alpha_{2000}$: 18$^{\rm h}$ 00$^{\rm m}$ -- 18$^{\rm h}$ 30$^{\rm m}$ and 
$\delta_{2000}$: -3$^{\circ}$ -- +2$^{\circ}$. Resulting distance extinction pairs. Median 
relation from Fig.~5 is overplotted. {\it Central panel}. Outcome of the Vilnius survey from 
Straizys et al. \cite{straizys02}.  {\it Bottom panel.} Extraction from Hipparcos. Distances are 
from van Leeuwen \cite{FVL07}, classification from Houk and Swift \cite{mich5}. For method see 
Knude and H{\o}g \cite{knude98}}\label{f3}
\end{figure}

The stars inside the dashed confinement of Fig.~\ref{f1} provide data for a first impression of the 
variation of extinction (e.g. $A_J$) vs. distance. Often an extinction rise is noticed as in 
Fig.~\ref{f3}. Some of the dwarfs inside the $(H-K)_0 \ vs. \ (J-H)_0$ standard relation but above 
the upper reddening vector in Fig.~\ref{f1} may possibly also be included by searching for stars 
that have extinctions and distances as indicated by the location and size of the $A_J$ jump. 
Assuming that they are $\sim$K dwarfs their absolute magitude is known from the $(J-K)_0 \ vs. \ M_J$ 
relation. In the case of Fig.~\ref{f3} distances between 150 and 250 pc and $A_J$ ranging from 
$\approx$0 to $\approx$1.2 mag outlines the parameter space where $\sim$K dwarfs are looked for. 
This addition to the sample might be small but the statistics for locating the $A_J$ jump by adding 
the 10 -- 20 K type dwarfs, that normally result, may be improved. But we emphasize that
the extinction -- distance study does not depend on the inclusion of these stars.

\section{Simple error progression}

For stars in the accepted parts of the CCD the intrinsic colors are estimated from translating
the observed position $(H-K, J-H)$ to the intrinsic main sequence position along a reddening
vector through the point. The intrinsic colors provide $(J-K)_0$ from which $M_J$ is estimated 
from the median relation shown in Fig.~\ref{f2}. The errors to be considered come from the 
observational errors and from the uncertainties in the slope and zero point in the reddening 
vector and in the two standard relations. The uncertainty in the intersection of the vector and 
the Bessell-Brett standard relation, approximated locally by a small line segment 
$(J-H)_0 \ = \ \alpha \times (H-K)_0 + \beta$, involves the uncertainty in the coefficients 
$\alpha$ and $\beta$. In a similar way we may estimate the error of the absolute magnitude $M_J$. 

The 2mass observations, the reddening ratio and the Bessell-Brett relation result in $(H-K)_0,
(J-H)_0$ and therefore in $(J-K)_0$ and e.g. $E(J-H)$ which with the adopted reddening law is
equivalent to $A_J$. Similarly the 2mass observations and the $(J-K)_0$ vs. $M_J$ relation 
provide the $M_J$ estimate. If $s=f(x_i)$ represents either the full set of equations used
to estimate $A_J$ or $M_J$ the formal error of $s$ follows from the progression formula:

\begin{equation}
\sigma_s^2 = \Sigma_i (\frac{\partial s}{\partial_{x_i}} \ \sigma_{x_i})^2
\end{equation}

For each star in the accepted part of the CCD we then compute $\frac{\partial s}{\partial_{x_i}}$
and  $\sigma_{x_i}$ implying that derivatives must be calculated and errors of the independent
parameters also must be known.

\begin{figure}
\epsfxsize=18.0cm
\epsfysize=18.0cm
\epsfbox{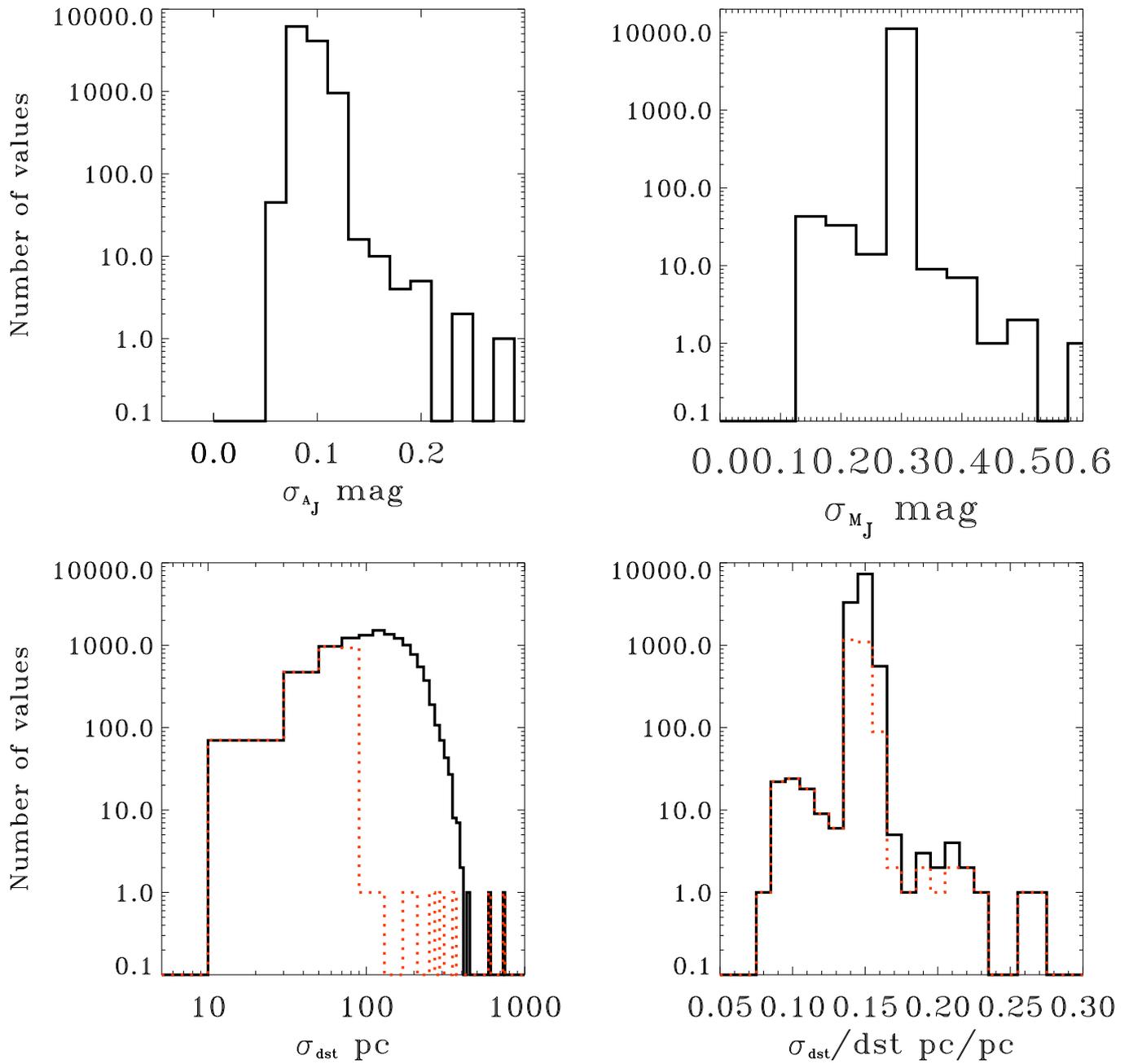} 
\caption[]{Distribution of errors for individual stars with $\sigma_{JHK} <$0.040 mag. Dashed 
curves for the two bottom diagrams pertain to stars with estimated distances less than 500 pc. 
The bottom right panel shows that the typical relative distance error is $\approx$15{\%} }\label{f4}
\end{figure}

The resulting error distributions for a sample, more than 12000 stars, in the 
Serpens -- Aquila region of the sky are shown in Fig.~\ref{f4}. 

But before we comment on the error distributions first a little on the sampled region. The Serpens 
star forming region has attracted much
attention over the years, e.g. Eiroa, Djupvik and Casali \cite{EDC08}, Straizys, Bartasiute and Cernis
\cite{straizys02} and Bontemps et al. \cite{bontemps10}. The area possesses active star forming 
regions and possibly contains sheets of interstellar material as well, Bontemps et al. 
\cite{bontemps10}.  In order to see if the estimated parameters are not completely off the mark
we have chosen to consider the rather large patch of sky studied by Straizys et al. 
\cite{straizys02} (see their Fig.~1), $\alpha_{2000}$: 18$^{\rm h}$ 00$^{\rm m}$ -- 18$^{\rm h}$ 
30$^{\rm m}$ and $\delta_{2000}$: -3$^{\circ}$ -- +2$^{\circ}$. 

In this region we have extracted all stars from 2mass with $\sigma_{JHK} < $ 0.040 mag with AAA
quality. After locating the stars in the region of acceptance of Fig.~\ref{f1}, we compute their 
extinction, distance and simultaneously do the error 
progression for the individual sources. The distance -- $A_J$ variation itself is displayed in
the upper panel of Fig.~\ref{f3} and the four panels of Fig.~\ref{f4} contain the distributions of
$\sigma_{A_J}, \ \sigma_{M_J}, \ \sigma_{distance}$ and most importantly the relative distance 
error $\sigma_{distance} / distance$ respectively. 

We may notice that despite quite a substantial number of main sequence stars are available in 
this part of the sky rather few are located within $\approx$200 pc. The upper left panel 
shows that for all these stars $A_J$ is estimated with an error 
better than $\sigma_{A_J}(max) \approx$ 0.2 mag corresponding to $\sigma_{A_V}(max) \ \lesssim$ 0.7 
mag. Only three stars have a larger error. The distribution is, however, rather narrow and peaks 
at $\lessapprox$0.1 mag corresponding to $\sigma_{A_V} \ \lessapprox$ 0.35 mag.

The distribution of the absolute magnitude errors is in the upper right panel. It has
a strongly peaked distribution with all stars having an almost identical error $\sigma_{M_J}
 \approx$ 0.3 mag. From Fig.~\ref{f2} we know that the early type stars have the largest uncertainty 
$\approx$0.4 mag and that the uncertainty becomes smaller, $\approx$0.1 mag, when the turn off is
approached. But most of the early type stars that may have been present in the 2mass Catalog 
are left out of the computations since their extinction, as mentioned, can not be derived with 
a sufficient accuracy. The uncertain intersection is caused by an almost parallelism of the 
reddening vector and the standard $(H-K)_0 \ - \ (J-H)_0$ relation for this spectral range.

Combining the extinction and luminosity errors the resulting distance error distribution comes
out as in the lower left panel of the Figure. Note the two logarithmic scales. The logarithmic 
scale is introduced on the x-axis in order to emphasize the smaller distances which have our greatest
attention. The dashed curve pertains to stars with an estimated distance less than 500 pc. The
total sample peaks at $\sigma_{distance} \approx$100 pc and the stars within 500 pc have errors
better than $\approx$80 pc. 

Finally we give the relative distance precision in the lower right of Fig.~\ref{f4}. The
majority of stars have 0.13 $<$ $\sigma_{distance} / distance$ $<$ 0.17 and we 
recall that this is the estimated relative error distribution for individual stars. 
The distribution peaks at $\approx$0.15.
An interesting point is that the relative distance error appears independent of distance when
$\sigma_{JHK} <$0.040 mag.
The dashed histogram are again for stars estimated to be within 500 pc. As a matter of
curiosity we mention that a relative precision of 15\% matches what was obtainable with
photoelectric $uvby\beta$ photometry for A and F stars brighter than V$\lesssim$9 mag a 
generation ago, Knude \cite{knude78}.

With a typical relative error of 15$\%$ on a single distance an uncertainty of $\sim$30 pc
for a star at 200 pc is implied.

\section{Extinction in the Serpens -- Aquila region} 

The Serpens -- Aquila region is known to contain several massive concentrations of
interstellar material whose distances are under debate. Bontemps et al. \cite{bontemps10} 
quotes a range of estimated distances to the clouds and star forming regions in this part of 
the sky. We do not enter this discussion presently but introduce the possibilty to use 
the 2mass -- Hipparcos calibration to estimate distances to such features. Fig.~\ref{f3} 
displays the distance -- $A_J$ variation as measured by individual stars. If one can show -- or
just assume --  that different stars in a common distance interval, e.g. 150 -- 180 pc are 
measuring extinctions pertaining to the same interstellar feature the error of the mean extinction 
and the mean distance - if that is relevant - may be calculated from the individual 
tracer errors. However, it has been a common pratice to identify an extinction 
discontinuity located at a fairly constant distance as caused by "an interstellar cloud". 
The solid curve in the three panels of Fig.~\ref{f3} is the median distance vs. median 
$J-band$ extinction calculated for 30 pc bins. If the extinction jump represents a cloud
its distance may be around 200 pc.

The central panel of Fig.~\ref{f3} is the outcome, with $A_V$ converted to $A_J$, of the Vilnius survey 
of the same region for which we have extracted 2mass data and the authors identified the 
distance to the front of clouds in this area to 225 $\pm$55 pc , Straizys et al. \cite{straizys02}.
For the sake of completeness we have included what may be obtained from Hipparcos stars
(HIP2: van Leeuwen \cite{FVL07}) and the Michigan spectral classification Vol. 5, Houk and 
Swift \cite{mich5} in the same area following Knude and H{\o}g \cite{knude98}.

In all three panels the median $A_J$ vs. distance relation is overplotted. If the distance to the 
rise of the median extinction is accepted as the cloud distance the cloud seems located somewhere
between 150 and 250 pc according to each of the three methods. 

The upper panel of Fig. \ref{f3} 
shows how the median extinction varies with the median distance. Each are calculated in 30 pc bins
separated by 15 pc steps. Bins within $\approx$200 pc are noticed to contain only a few stars and the 
error of the mean will match the individual errors rather closely. For larger distances the error 
of the mean extinction and distance will benefit from the larger number of stars in a given 
distance bin, see Table~\ref{t1}. In Fig. \ref{f5} the two top panels show the median variation 
and the corresponding errors of the mean(s). The center panel has a logarithmic distance scale 
to stress distances closer than $\approx$300 pc. The bottom panel of the diagram shows how the 
average line of sight density $A_J / distance \ \propto \ n_H$ atoms cm$^{\rm -3}$ formed from the
median values vary with distance. 
A density peak, statistical significant according the errors of the mean, see Table~\ref{t1} for the
numbers displayed on the Figure, 
is present at $\approx$200 pc. If we define the distance of the cloud as the distance with the 
maximum average line of sight density $n_H$ we have an indication from the 
$A_J / distance \ vs. \ distance$ diagram and even an estimate of the error on this distance: 
203 $\pm$ 7 pc. The estimated error $\pm$7 pc is based on the individual errors pertaining to the 177
stars located in the 30 pc wide bin with the median distance 203 pc. In a recent report using a different method we estimated the distance to the 
Serpens cloud as 193 $\pm$ 13 pc, Knude \cite{knude10}. What we do not known is whether the 
extinctions and distances we estimate can be attributed to the star forming Serpens clouds -- only that 
some material appears in the general direction of the clouds.        

\section{Errors in mean extinction and distance} 
\begin{figure}
\epsfxsize=17.0cm
\epsfysize=21.0cm
\epsfbox{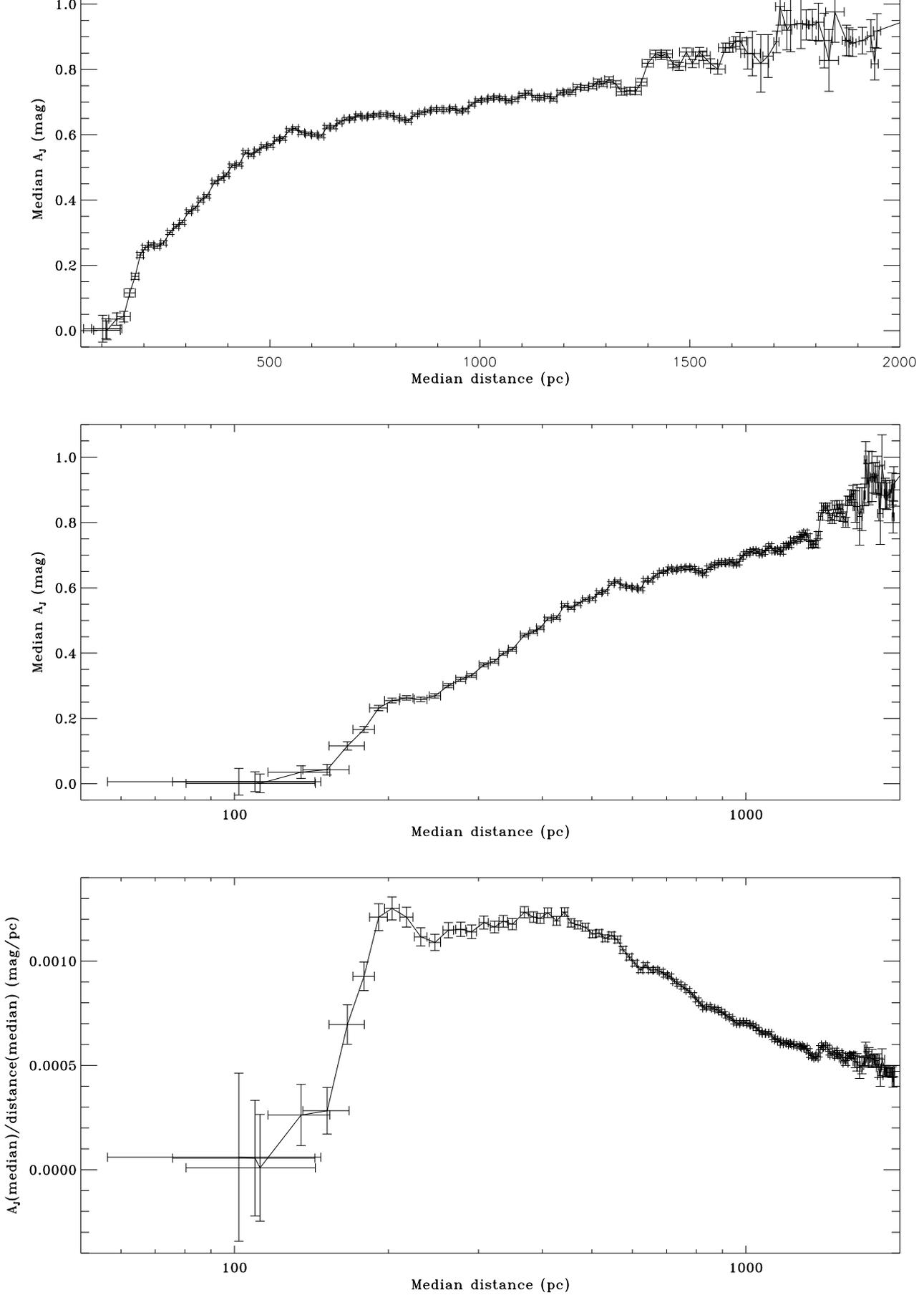} 
\caption[]{{\it Upper panel.} Median of distances and extinctions $A_J$ calculated for 30 pc bins.
The median points are plotted together with the error of the mean also for 30 pc bins. {\it Middle panel.}
Same as {\it Upper panel} but on a logarithmic distance scale to emphasize nearby bins. 15 nearest bins
are tabulated in Table \ref{t1}.
{\it Bottom panel} Line of sight average density $A_J / distance$ with errors}\label{f5}
\end{figure}

If the extinction of all stars in a given distance range measure the same interstellar feature,
which probably will have a column density variation from the cloud rim, where it approximates what
is caused by the general diffuse interstellar medium, to dense interior cores, we may use 
median values to indicate a distance estimate to this feature. In Knude \cite{knude10} it was suggested
that $\overline{H-K}_{reseau}$ contours, where the reseau size was defined from the requirement that a
reseau should hold 100 stars on the average, was useful to define cloud outlines.  
Note that because our sample is only selected from the region of acceptance of Fig.~\ref{f1} large
extinctions are not measured:  $A_J(max)$ $\lesssim$1.5 mag. As the upper panel of Fig.~\ref{f3} 
shows the $A_J$ increases around 200 pc. Noting that the relative distance error is about 15$\%$ 
we have bined the stars in 30 pc intervals. In Table~\ref{t1} we show values for the first 15 
distance intervals and results are plotted in Fig.~\ref{f5}. The upper panel displays the variation 
of the medians: $distance(median) \ vs. \ A_J(median)$. For each distance bin the error of the 
mean for both variables are also displayed. The central panel has a logarithmic scale to emphasize 
the nearer distances but is otherwise identical to the upper panel. The lower panel indicates that 
cloud distances possibly may be identified by a local maximum in the line of sight average density 
$A_J / distance$ mag/pc $\propto$ \ $n_H$ atoms/cm$^{\rm 3}$. Errors of distance as well as the 
average line of sight density are indicated.
 
\begin{center}
\begin{table}
\caption{Median distance and extinction in the $J-band$ band together with errors of the mean.
The mean and median are calculated for 30 pc distance bins. Column 5 indicates the
number of stars in a bin. Column 6 is $A_J(median)/distance(median)$ in mag/pc and Column 7 is the
error of this ratio based on the errors of Column 2 and 4. Exclusively from stars in the region of
acceptance and the minor set of secondary K dwarfs all with $\sigma_{JHK}<$0.040 mag}
\hspace{0.88cm}

\begin{tabular}{|c|r|c|r|r|r|r|}
\hline
Distance&Error&$A_J$&Error&{\#}&$A_J/distance$&$\sigma (A_J/distance)$\\
\hline
pc & pc&mag&mag& &mag/pc&mag/pc \\
\hline
 & & & & & &\\
102.0& 45.5&0.0061&0.0410&6&5.98039e-05&0.000403061\\
109.7& 34.0&0.0061&0.0303&11&5.56062e-05&0.000277036\\
112.2& 31.7&0.0010&0.0287&12&8.91266e-06&0.000256118\\
135.0& 18.6&0.0354&0.0192&26&0.000262222&0.000146937\\
151.8& 15.6&0.0429&0.0163&38&0.000282609&0.000111483\\
166.3& 13.1&0.1157&0.0127&62&0.000695731&9.46005e-05\\
179.2& 8.6&0.1661&0.0092&113&0.000926897&6.83839e-05\\
191.4& 7.6&0.2317&0.0081&143&0.00121055&6.42917e-05\\
203.3& 6.7&0.2547&0.0073&177&0.00125283&5.51005e-05\\
217.0& 6.3&0.2627&0.0067&206&0.00121060&4.74362e-05\\
231.4& 6.6&0.2583&0.0068&204&0.00111625&4.35682e-05\\
246.7& 6.3&0.2688&0.0065&221&0.00108958&3.86119e-05\\
262.0& 6.2&0.3008&0.0063&239&0.00114809&3.66818e-05\\
277.1& 6.1&0.3192&0.0062&246&0.00115211&3.41028e-05\\
290.9& 6.3&0.3317&0.0061&256&0.00114025&3.25962e-05\\
 & & & & & &\\
\hline
\end{tabular}
\label{t1}
\end{table}
\end{center}

\subsection{Another Cloud Distance Indicator?}

\begin{figure}
\epsfxsize=18.0cm
\epsfysize=18.0cm
\epsfbox{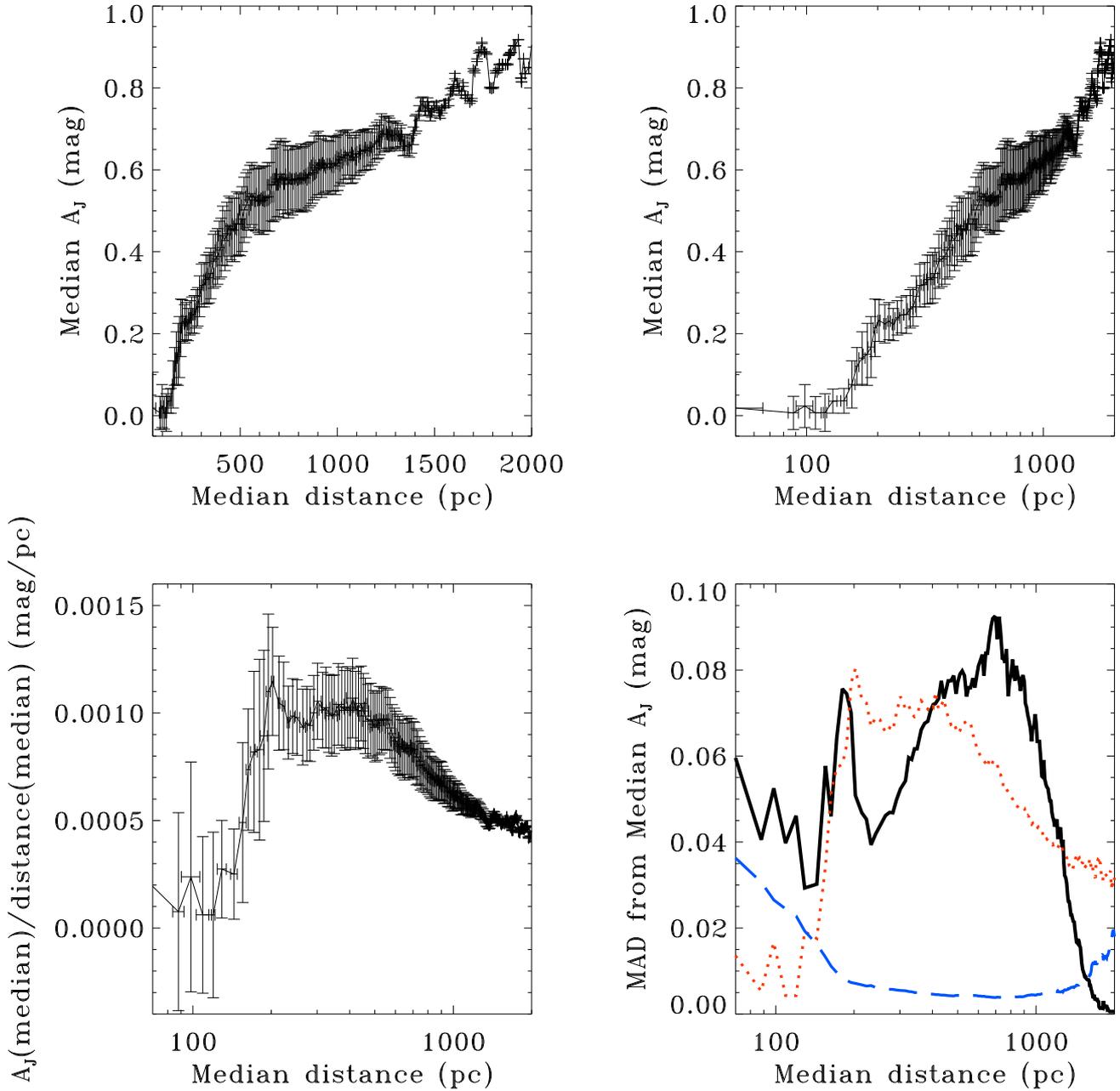} 
\caption[]{Set of diagrams including MAD: Median Absolute Deviation from the median, considered to
be a rather robust statistics. {\it Upper left.} Median curve with error bars indicating the Median 
Absolute Deviation of distance and extinction. {\it Upper right.} Same as {\it Upper left} but on a 
logarithmic distance scale. {\it Lower left.} Average density $A_J(median)$/$Distance(median)$. MAD of both
coordinates is plotted. {\it Lower right.} MAD($A_J)$ vs. median distance is plotted as the solid curve. 
The dotted curve is a scaled version of the average density (multiplied by 70). The dashed curve is the 
run of the error of the mean extinction vs. distance. From Fig. \ref{f4} we recall that the typical error 
of $A_J$ is $\lesssim$0.1 mag. The near coincidence of MAD($A_J)$ and the peak of
the average line of sight density is noticed}\label{f6}
\end{figure}

Traditionally the location of a cloud has been identified by the presence of an extinction jump at a 
rather well defined distance and the distance(s) of the nearest star(s) showing the extinction rise have 
been used to estimate a lower limit to the cloud distance. Star forming clouds at least are expected
to show a large variation in column densities and one could imagine that a cloud do have a distribution
of deviations from the median column density different from that pertaining to stars at a given distance
and situated outside clouds. A possible statistics able to show the signature of a cloud could be the 
Median Absolute Deviation from the median of $A_J$, MAD($A_J$). In Fig. \ref{f6} we show a preliminary
presentation of the possible use of MAD($A_J$). In the {lower ight} panel is shown the variation of
MAD($A_J$) with median distance and the near coincidence of an extremum of MAD($A_J$) and the peak in
the line of sight average density $A_J(median)$/$Distance(median)$ is noticed.  

\section{Acknowledgements}
This publication makes use of data products from the Two Micron All
Sky Survey, which is a joint project of the University of Massachusetts
and the Infrared Processing and Analysis Center/California Institute of
Technology, funded by the National Aeronautics and Space Administration
and the National Science Foundation. This research has made use of the SIMBAD
database, operated at CDS, Strasbourg, France.

\end{document}